\documentclass[aps,prd,groupedaddress,showpacs,nofootinbib]{revtex4}

\usepackage{graphicx,dcolumn,bm,amssymb,amsmath,latexsym,amsfonts,footnote}

\begin{document}

\newcommand{\nonu}{\nonumber}
\newcommand{\sm}{\small}
\newcommand{\noi}{\noindent}
\newcommand{\npg}{\newpage}
\newcommand{\nl}{\newline}
\newcommand{\bp}{\begin{picture}}
\newcommand{\ep}{\end{picture}}
\newcommand{\bc}{\begin{center}}
\newcommand{\ec}{\end{center}}
\newcommand{\be}{\begin{equation}}
\newcommand{\ee}{\end{equation}}
\newcommand{\beal}{\begin{align}}
\newcommand{\eeal}{\end{align}}
\newcommand{\bea}{\begin{eqnarray}}
\newcommand{\eea}{\end{eqnarray}}
\newcommand{\bnabla}{\mbox{\boldmath $\nabla$}}
\newcommand{\univec}{\textbf{a}}
\newcommand{\VectorA}{\textbf{A}}
\newcommand{\Pint}

\title{Metric for two unequal extreme Kerr-Newman black holes}

\author{I. Cabrera-Munguia\footnote{icabreramunguia@gmail.com}}
\affiliation{ Departamento de F\'isica y Matem\'aticas, Universidad Aut\'onoma de Ciudad Ju\'arez, 32310 Ciudad Ju\'arez, Chihuahua, M\'exico}


\begin{abstract}
In the present paper, within the framework of stationary axisymmetric spacetimes, binary systems composed of two unequal co- and counter-rotating extreme Kerr-Newman black holes separated by a massless strut are reported. The metric describing both configurations is introduced in a closed analytical form in terms of five arbitrary parameters: the masses $M_{i}$, electric charges $Q_{i}$, and a coordinate distance $R$. We obtain novel results from these configurations; in particular, those related to the merging process.
\end{abstract}

\maketitle

\vspace{-0.4cm}
\section{Introduction}
\vspace{-0.3cm}
Binary black hole (BH) systems have attracted our attention since the early days of general relativity. The recent detection of gravitational waves \cite{LIGO} produced by binary BH mergers permits us to reconsider exact binary models to complement the vast amount of numerical results in the literature. However, from a technical point of view, it is quite complicated to take into account all the dynamical interactions in a binary setup, and for such a reason the stationary scenario seems to be a good candidate to develop analytical results. In static charged systems, the Majumdar-Papetrou metric \cite{Majumdar,Papapetrou,HH} describes the simplest model of two extreme BHs, which remain in neutral equilibrium due to the balance of their electric charges and masses according to the relation $Q_{i}=\pm M_{i}$, regardless of the separation distance among sources. Moreover, in vacuum systems, the Kinnersley-Chitre (KCH) exact solution \cite{KCH} allows us the description of rotating binary BHs, after solving appositely the axis conditions \cite{MR,ICM2018}. In this type of binary vacuum systems, the Kerr BHs are apart by a conical singularity \cite{BachW,Israel}, which can give us information on their gravitational attraction and spin-spin interaction.

In contrast, the treatment of unequal binary configurations of extreme Kerr-Newman (KN) BHs \cite{ENewman} has been a fairly complicated problem beyond our possibilities, due mainly to the fact that the axis conditions are not enough to define properly KN BHs, therefore, it is necessary to impose an extra condition in order to kill both magnetic charges, otherwise, Dirac strings linked to the KN BHs will appear \cite{Tomi, Galtsov, ICM2020}. The main purpose of this paper is to derive a five-parametric exact solution that completely describes binary co- and counter-rotating extreme KN BH separated by a massless strut in a unified manner. To accomplish such a goal, we are going to take into account the recent results of \cite{ICM2021} where a complete derivation of the metric and thermodynamical properties for non-extreme KN BHs has been succeeded. Hence, the Ernst potentials and metric functions will be depicted in terms of physical Komar parameters \cite{Komar}: the masses $M_{i}$, electric charges $Q_{i}$, and a coordinate distance $R$ as well. In this scheme, the five arbitrary parameters compose an algebraic equation thus defining a dynamical law for interacting BHs with struts, which is reduced to some previous studied cases \cite{ICM2018, ICM2015}. At the same time, the metric is concisely given in terms of Perjes' factor structure \cite{Perjes}. Since the physical limits in both rotating charged models are well identified, after turning our sight exclusively in the corotating binary BH setup, we derive quite simple formulas for the area of the horizon and the interaction force during the merger limit. In addition, a deformed metric for a near horizon extreme binary KN BH is also given.

\vspace{-0.4cm}
\section{The charged Kinnersley-Chitre exact solution}
\vspace{-0.3cm}
Ernst's formalism \cite{Ernst} allows the description of Einstein-Maxwell equations in stationary axisymmetric spacetimes, in terms of a pair of complex functions $({\cal{E}}, \Phi)$ satisfying
\vspace{-0.1cm}
\bea \begin{split}  \left({\rm{Re}} {\cal{E}}+|\Phi|^{2}\right)\Delta{\cal{E}}&=(\bnabla{\cal{E}}+
2\bar{\Phi}\bnabla \Phi)\cdot\bnabla {\cal{E}}, \\
\left({\rm{Re}}{\cal{E}}+|\Phi|^{2}\right)\Delta \Phi&=(\bnabla{\cal{E}}+
2\bar{\Phi}\bnabla\Phi)\cdot \bnabla\Phi. \label{ERNST} \end{split} \eea

\vspace{-0.1cm}
\noi where any exact solution of Eq.\ (\ref{ERNST}) can be derived via Sibgatullin's method (SM) \cite{Sibgatullin,MSO}, which is also useful to obtain the metric functions $f(\rho,z)$, $\omega(\rho,z)$ and $\gamma(\rho,z)$ of the line element \cite{Papapetrou}
\vspace{-0.1cm}
\be ds^{2}=f^{-1}\left[e^{2\gamma}(d\rho^{2}+dz^{2})+\rho^{2}d\varphi^{2}\right]- f(dt-\omega d\varphi)^{2}.
\label{Papapetrou}\ee

\vspace{-0.1cm}
Due to the fact that SM needs a particular form of the Ernst potentials on the upper part of the symmetry axis, let us begin with a more suitable physical representation, namely
\vspace{-0.1cm}
\begin{align}
{\cal E}(0,z)&=\frac{\mathfrak{e}_{1}}{\mathfrak{e}_{2}}, \qquad \Phi(0,z)=\frac{\mathcal{Q}z+\mathfrak{q}_{o}}{\mathfrak{e}_{2}}, \nonu \\
\mathfrak{e}_{1}&=z^{2}-[M + i(\mathfrak{q}+2J_{0})]z +P_{+}+i P_{1} -2iJ_{0}(d-i\mathfrak{q}), \qquad
\mathfrak{e}_{2}=z^{2} + (M -i\mathfrak{q})z + P_{-} + i P_{2}, \nonu\\
P_{\pm}&= \frac{M(2\Delta_{o}-R^{2})\pm 2\left[\mathfrak{q}{\rm s}_{1}-2(q_{o}Q+b_{o} B)\right]}{4M},\qquad
{\rm s}_{1}=P_{1}+P_{2}, \qquad d=M+\frac{P_{2}}{\mathfrak{q}},\qquad \mathcal{Q}=Q+iB, \nonu\\
\mathfrak{q}_{o}&=q_{o}+ib_{o}, \qquad \Delta_{o}= M^{2}-|\mathcal{Q}|^{2}-\mathfrak{q}^{2},
 \label{ernstaxiselectro}\end{align}

\vspace{-0.1cm}
\noi where the aforementioned Ernst potentials Eq.\ (\ref{ernstaxiselectro}) are the extreme case of that one considered in Ref.\ \cite{ICM2021}, from which the first Simon's multipole moments \cite{Simon} can be explicitly calculated by means of the Hoenselaers-Perj\'es procedure \cite{HP,Sotiriou}. In this sense, $M$ plays the role of the total mass of the system and $Q+iB$ defines the total electromagnetic charge, while the total electric and magnetic dipole moment are $q_{o}-B(\mathfrak{q}+J_{0})$ and $b_{o}+Q(\mathfrak{q}+J_{0})$, respectively. Besides, $R$ represents a separation distance between both sources. At the same time, the NUT charge $J_{0}$ \cite{NUT} and total angular momentum of the system $J$ are given by
\vspace{-0.1cm}
\begin{align}
J_{0}&=\frac{N}{8M^{2}(\mathfrak{q}P_{-}+P_{2}d)} , \qquad J=M\mathfrak{q}-\frac{{\rm s}_{2}}{2}+(M+d)J_{0}, \nonu\\
N&=M^{2}\left\{4(P_{1}P_{2}+|\mathfrak{q}_{o}|^{2})-\Delta_{o}(R^{2}-\Delta_{o})\right\} 
-\left[\mathfrak{q}{\rm s}_{1}-2(Q q_{o}+ B b_{o})\right]^{2}, \qquad
{\rm s}_{2}=P_{1}-P_{2},\label{Multipolarterms}\end{align}

\vspace{-0.1cm}
It is not difficult to show that once the SM \cite{Sibgatullin,MSO} has been applied to the axis data Eq.\ (\ref{ernstaxiselectro}), the Ernst potentials satisfying Eq.\ (\ref{ERNST}) acquire the final aspect
\vspace{-0.1cm}
\begin{align} {\cal{E}}&=\frac{\Lambda-2\Gamma}{\Lambda+2\Gamma},\qquad \Phi=\frac{2 \chi}{\Lambda+2\Gamma}, \nonu\\
\Lambda&=R^{2}\left[(R^{2}-\delta)(x^{2}-y^{2})^{2}+\delta(x^{4}-1)\right] +\Big\{|\mathfrak{p}|^{2}+(\mathfrak{q}+J_{0}) \mathfrak{r} -R^{2}(R^{2}-\delta)\Big\}(y^{4}-1)\nonu\\
&+2iR \Big\{xy\Big[\big[\mathfrak{r}+(\mathfrak{q}+J_{0})R^{2}\big](y^{2}-1)-(\mathfrak{q}+J_{0})R^{2}(x^{2}+y^{2}-2)\Big] -R{\rm S}_{1}\big(x^{2}+y^{2}-2x^{2}y^{2}\big) \Big\} \nonu\\
\Gamma&=(M+iJ_{0})\mathbb{P}_{1}-(\mathfrak{b}+i{\rm S}_{2})\mathbb{P}_{2}, \qquad
\chi= \mathcal{Q} \mathbb{P}_{1}+2\mathfrak{q}_{o} \mathbb{P}_{2}\qquad
\mathbb{P}_{1}=R^{3} x(x^{2}-1)-(R \bar{\mathfrak{p}} x-i \mathfrak{r} y)(y^{2}-1),\nonu\\
\mathbb{P}_{2}&= R^{2}y (x^{2}-1)-\big[\mathfrak{p}y-i(\mathfrak{q}+J_{0})Rx\big](y^{2}-1),\qquad
\mathfrak{p}=R^{2}-\delta+i{\rm S}_{1},\qquad \mathfrak{r}=2\mathfrak{a}-(\mathfrak{q}+J_{0})(R^{2}-2\delta)-2\mathfrak{b}J_{0}, \nonu\\ 
\mathfrak{a}&=M{\rm S}_{2}+2(b_{o}Q-q_{o}B), \qquad \delta=\Delta_{o}-2\mathfrak{q}J_{0}, \qquad
{\rm S}_{1}={\rm s}_{1}-2d J_{0}, \qquad
{\rm S}_{2}={\rm s}_{2}-2d J_{0}, \nonu\\
\mathfrak{b}&=\big[\mathfrak{q}{\rm s}_{1}-2(q_{o}Q+b_{o}B)\big]/M-2\mathfrak{q}J_{0},
\label{Ernstextreme}\end{align}

\vspace{-0.1cm}
\noi where $(x,y)$ are prolate spheroidal coordinates related to cylindrical coordinates $(\rho,z)$ by means of \vspace{-0.1cm}
\be x=\frac{r_{+}+r_{-}}{R}, \quad y=\frac{r_{+}-r_{-}}{R}, \quad r_{\pm}=\sqrt{\rho^{2} + (z \pm R/2)^{2}}.   \label{prolates}\ee

\vspace{-0.1cm}
Furthermore, the metric functions contained within the line element can be written down in a closed analytical form by using Perjes's factor structure \cite{Perjes}, thus getting\footnote{It is possible to locate up or down  semi-infinite singularities along the axis depending on the values for $C=0, \pm 1$ (see \cite{MR} and references therein)}
\vspace{-0.1cm}
\begin{align} f&=\frac{\mathcal{D}}{\mathcal{N}},\qquad \omega=2J_{0}(y+C)+\frac{R(y^{2}-1) \big[(x^{2}-1)\Sigma\Pi-\Theta {\rm T}\big]}{2\mathcal{D}}, \qquad
e^{2\gamma}=\frac{\mathcal{D}}{R^{8}(x^{2}-y^{2})^{4}}, \nonu\\
\mathcal{N}&= \mathcal{D}+ \Theta \Pi-(1-y^{2})\Sigma {\rm T}, \quad
\mathcal{D}= \Theta^{2}+(x^{2}-1)(y^{2}-1)\Sigma^{2}, \nonu\\
\Theta&=R^{2}\Big[(R^{2}-\delta)(x^{2}-y^{2})^{2}+\delta(x^{2}-1)^{2}\Big]
+\Big[|\mathfrak{p}|^{2}+(\mathfrak{q}+J_{0}) \mathfrak{r}
-R^{2}(R^{2}-\delta) \Big](y^{2}-1)^{2}, \nonu\\
\Sigma&=2R\Big((\mathfrak{q}+J_{0})R^{2}x^{2}-\mathfrak{r}y^{2} -2R{\rm S}_{1}xy\Big), \nonu\\
\Pi&= 4Rx\bigg\{MR^{2}(x^{2}-y^{2})+\big[M\delta+(\mathfrak{q}+2J_{0}){\rm S}_{2}+2J_{0}{\rm S}_{1}\big]
 (1+y^{2})+(2M^{2}+2J_{0}^{2}-|\mathcal{Q}|^{2})Rx -2{\rm S}_{1}\big[\mathfrak{q}y+2J_{0}(1+y)\big]\bigg\}\nonu\\
&\hspace{-0.5cm} -4y\bigg\{ \mathfrak{b}\Big(R^{2}(x^{2}-y^{2})+\delta(1+y^{2})+2MRx-2\mathfrak{b}y\Big) +{\rm S}_{1}{\rm S}_{2}
(1+y^{2}) -2\big({\rm S}_{2}^{2}-2|\mathfrak{q}_{o}|^{2}\big)y +J_{0}
\big[\mathfrak{r}(1-y^{2})+2(\mathfrak{q}+J_{0})R^{2}x^{2}\big] \bigg\}, \nonu
\end{align}

\vspace{-0.1cm}
\begin{align} 
{\rm T}&=\frac{2}{R} \bigg\{2R^{2}\Big(\big[M{\rm S}_{1}-(\mathfrak{q}+J_{0})\mathfrak{b}+J_{0}(R^{2}-\delta) \big]y
-{\rm S}_{2}(Rx+M) -\mathfrak{a}+2\mathfrak{b}J_{0}\Big)(1-x^{2})\nonu\\
&+\bigg(2\big[\mathfrak{b}{\rm S}_{1}+(R^{2}-\delta){\rm S}_{2}+M \mathfrak{r}\big](Rx+M) + 2J_{0}\Big[{\rm S}_{1}{\rm S}_{2}-\mathfrak{b}(2R^{2}-\delta)- \big[|\mathfrak{p}|^{2}+(\mathfrak{q}+J_{0})\mathfrak{r}\big]y\Big]+2\mathfrak{a} R^{2} -(M^{2}-\mathfrak{q}^{2})\mathfrak{r} \nonu\\
&+(\mathfrak{q}+J_{0})\Big[\delta R^{2}+2J_{0}\mathfrak{r}-4|\mathfrak{q}_{o}|^{2}\Big]\bigg)
(1-y^{2})\bigg\}.\label{extreme}\end{align}

\vspace{-0.1cm}
The above metric is the electromagnetically charged version of KCH's exact solution \cite{KCH,MR}. It contains nine parameters defined by the set $\{M,R,\mathfrak{q},P_{1},P_{2},Q,B,q_{o},b_{o}\}$. In the absence of electromagnetic field ($\mathcal{Q}$ and $\mathfrak{q}_{o}$ set to zero) the KCH exact solution \cite{KCH, MR} emerges straightforwardly after doing the simple redefinitions \cite{ICM2018}
\vspace{-0.1cm}
\begin{align}
M&=\frac{2\kappa(p\mathrm{P}-p\mathrm{Q}\alpha+q\mathrm{P}\beta)}{p^{2}+\alpha^{2}-\beta^{2}},\qquad
J_{0}=-\frac{2\kappa(p\mathrm{Q}+p\mathrm{P}\alpha+q\mathrm{Q}\beta)}{p^{2}+\alpha^{2}-\beta^{2}},\nonu\\
P_{1}&=\frac{2\kappa^{2}[(1-q\mathrm{Q})\alpha-p\mathrm{P}\beta]}{p^{2}+\alpha^{2}-\beta^{2}}+
2dJ_{0}, \qquad
P_{2}= \frac{2\kappa^{2}[(1+q\mathrm{Q})\alpha+p\mathrm{P}\beta]}{p^{2}+\alpha^{2}-\beta^{2}}, \qquad e^{-i\gamma_{o}}=\mathrm{P}-i \mathrm{Q}, \nonu \\
\mathfrak{q}&=\frac{2\kappa[p(q+\mathrm{Q})+ p \mathrm{P} \alpha +(1+q\mathrm{Q})\beta]}{p^{2}+\alpha^{2}-\beta^{2}},  \qquad R=2\kappa, \label{relations1}\end{align}

\vspace{-0.1cm}
\noi where $p^{2}+q^{2}=1$ and $|e^{-i\gamma_{o}}|=1$. Taking into account Bonnor's description \cite{Bonnor}, the above metric is not asymptotically flat due to the presence of the NUT charge which represents a semi-infinite singular source located along the lower part of the symmetry axis at $y=-1$, for $C=-1$, thus providing additional rotation to the binary system. The last point can be better understood when analyzing the asymptotic behavior of the metric functions; i.e., $f \rightarrow 1$, $\gamma \rightarrow 0$, and $\omega \rightarrow 2J_{0}(y+C)$ at $x \rightarrow \infty$. In this regard, the condition $J_{0}=0$ is enough to ensure any asymptotically flat spacetime from Eq.\ (\ref{extreme}). Such a task can be accomplished by means of
\vspace{-0.1cm}
\begin{align}
&\hspace{-0.3cm} M^{2}\left[4(P_{1}P_{2}+|\mathfrak{q}_{o}|^{2})-\Delta(R^{2}-\Delta)\right]
-\left(\mathfrak{q}{\rm s}_{1}-2Q q_{o}\right)^{2}=0, \nonu\\
\Delta&=M^{2}-Q^{2}-\mathfrak{q}^{2}, \label{noNUT}\end{align}

\vspace{-0.1cm}
\noi where we have imposed also the requirement $B=0$ in order to describe afterward extreme KN binary BHs. In addition, the condition $\omega(x=1,y=2z/R)=0$ permitting to disconnect the region among the sources is  reduced to
\vspace{-0.2cm}
\begin{align}
&2(R+M)\bigg\{\Big[\mathfrak{q}{\rm s}_{1}-2q_{o}Q\Big]{\rm s}_{1}+M(R^{2}-\Delta){\rm s}_{2} -\mathfrak{q}M^{2}R^{2}\bigg\}+2MP_{0}\Big[M{\rm s}_{2}+2b_{o}Q+\mathfrak{q} \Delta \Big] +M\mathfrak{q} (Q^{2}R^{2}-4|\mathfrak{q}_{o}|^{2})=0, \nonu\\
P_{0}&=(R+M)^{2}+\mathfrak{q}^{2}.\label{disconnect}\end{align}

\vspace{-0.1cm}
If one is able to get an analytical solution from Eqs.\ (\ref{noNUT}) and (\ref{disconnect}), it will be possible to derive a binary model of rotating dyonic extreme BHs kept apart by a massless strut, where the sources are equipped with identical magnetic charges but endowed with opposite signs. Therefore, there exists a Dirac string joined to the BHs unless the magnetic charges are removed from the solution \cite{Tomi, Galtsov, ICM2020}. In this case, an absence of individual magnetic charges is accomplished if the condition on the real part of the potential $\Phi$ is imposed \cite{Tomi}, that is
\vspace{-0.1cm}
\be \lim_{\lambda \rightarrow 0} \Big[{\rm Re}\Phi\big(x=1+\lambda,y=1\big)-{\rm Re}\Phi\big(x=1,y=1-\lambda\big)\Big]=0, \label{nomagneticcharge}\ee

\vspace{-0.1cm}
\noi and this condition only is established in the upper BH since the lower one contains the same magnetic charge with opposite sign as we have before mentioned. A straightforward calculation yields the following expression
\vspace{-0.1cm}
\begin{align}
&\mathfrak{q}Q\bigg\{ \Big[ MP_{0}-(R+M)(2M(R+M)-\mathfrak{q}^{2})\Big]{\rm s}_{1}^{2} -4M^{2}
(R+M)P_{1}P_{2}\bigg\} +M^{2}\Big[2\mathfrak{q}P_{0}b_{o}-Q\big(P_{0}-2\mathfrak{q}^{2}\big)
(R^{2}-\Delta) \Big]{\rm s}_{2} \nonu\\
&-2q_{o}\Big[ M(\Delta+M R)P_{0}+2\mathfrak{q}^{2}(R+2M)Q^{2}\Big]{\rm s}_{1} -4\mathfrak{q}q_{o}^{2}Q\left[ MP_{0}-(R+M)Q^{2}\right] \nonu\\
&-M^{2}(R+M)(R^{2}-\Delta) \left[ 2P_{0}b_{o}+\mathfrak{q} Q(R^{2}-\Delta) \right]=0,  \label{noBcharge}
\end{align}

\vspace{-0.1cm}
\noi where this algebraic equation killing both magnetic charges has recently been got in Ref.\ \cite{ICM2021} for the case of non-extreme sources. The most general exact solution cannot be derived directly from these entwined
equations unless we get first a highly complicated fourth-degree algebraic equation, consequently, one must circumvent this technical issue by adopting a different point of view. As we shall see next, we are going to derive the algebraic values for the set $\{q_{o},b_{o},P_{1},P_{2}\}$ that defines extreme KN binary BHs held apart by a massless strut in a physical representation.

\vspace{-0.4cm}
\section{Extreme KN binary BHs}
\vspace{-0.3cm}
Let us begin the section by considering first the variables $\{q_{o},b_{o},P_{1},P_{2}\}$ earlier derived in Ref.\ \cite{ICM2021} and containing a physical representation are given by
\vspace{-0.1cm}
\begin{align}
q_{o}&=\frac{2\mathfrak{q} (Q_{1}J_{2}-Q_{2}J_{1})}{P_{0}}+\frac{Q_{1}}{2} (R-2M_{2})-\frac{Q_{2}}{2} (R-2M_{1}),\nonu\\
b_{o}&=\Bigg[(Q_{1}C_{2}-Q_{2}C_{1})\bigg(\frac{J_{1}}{\mathcal{P}_{1}}-\frac{J_{2}}{\mathcal{P}_{2}}\bigg)
-\frac{\mathfrak{q}}{P_{0}}\Bigg(Q - \frac{Q_{1}H_{1+}\big[Q_{1}(Q_{1}-Q_{2})P_{0}-2(R+M)H_{1-}\big]}{\mathcal{P}_{1}} \nonu\\
&-\frac{Q_{2}H_{2+}\big[Q_{2}(Q_{2}-Q_{1})P_{0}-2(R+M)H_{2-}\big]}{\mathcal{P}_{2}} \Bigg) \Bigg] \frac{(R^{2}-\Delta)}{2}, \nonu\\
P_{1,2}&=\frac{(2H_{2}A_{2}-RP_{0}\mathcal{P}_{2})J_{1}-(2H_{1}A_{1}-RP_{0}\mathcal{P}_{1})J_{2}}
{2P_{0}\mathcal{P}_{0}}\pm (M\mathfrak{q}-J),
\label{simplePxandQx}\end{align}

\vspace{-0.1cm}
\noi with the following elements:
\vspace{-0.1cm}
\begin{align}
A_{i}&=\mathcal{P}_{i}-(R^{2}-\Delta)H_{i+}P_{0}, \qquad H_{i}=M_{i}P_{0}-Q_{i}Q(R+M),\qquad
\mathcal{P}_{i}=H_{i-}C_{i}-(-1)^{i}(M_{1}-M_{2})Q^{2}_{i}P_{0}^{2},\nonu\\
\mathcal{P}_{0}&=M\mathcal{P}_{1}-(R^{2}-\Delta)H_{1+}H_{1-}\equiv M\mathcal{P}_{2}-(R^{2}-\Delta)H_{2+}H_{2-}, \qquad
C_{i}=P_{0}^{2}-2M_{i}(R+M)P_{0}+2\mathfrak{q}^{2} Q_{1}Q_{2}, \nonu\\
H_{i\pm}&=M_{i}P_{0}\pm Q_{1}Q_{2}(R+M), \qquad i=1,2.
\label{elements}\end{align}

\vspace{-0.1cm}
Also, in the binary setup, the BH horizons $\sigma_{i}$ are expressed as \cite{ICM2021}
\vspace{-0.1cm}
\begin{align}
\sigma_{i}&=\sqrt{D_{i}-J_{i}\left( \frac{J_{i}G_{i}-2\mathfrak{q}A_{i}B_{i}}{P_{0}^{2}\mathcal{P}_{i}^{2}}\right)},\nonu\\
D_{i}&=M_{i}^{2}- Q_{i}^{2}F_{i}-2(-1)^{i}Q_{i}F_{0},\qquad
G_{i}=\left[2(R+M)\mathcal{P}_{i}+P_{0}(R^{2}-\Delta)C_{i}\right]^{2}-4P_{0}\mathcal{P}_{1}\mathcal{P}_{2},\nonu\\
F_{0}&=\frac{M_{2}Q_{1}-M_{1}Q_{2}}{R+M}\left( 1-\frac{\mathfrak{q}^{2}}{P_{0}^{2}}\right),\qquad  F_{i}=1-\frac{Q_{i}^{2}\mathfrak{q}^{2}}{P_{0}^{2}}
\left(1-\frac{A_{i}^{2}}{\mathcal{P}_{i}^{2}}\right)+\frac{Q^{2}\mathfrak{q}^{2}}{P_{0}^{2}},\qquad
B_{i}=Q_{i}^{2}P_{0}(R^{2}-\Delta)C_{i}-2H_{i}\mathcal{P}_{i}, \nonu\\ 
i&=1,2,\label{sigmas}\end{align}

\vspace{-0.1cm}
\noi where the set of parameters $\{M_{1},M_{2},Q_{1},Q_{2},J_{1},J_{2}\}$ are the physical Komar parameters \cite{Komar} for each source. It should be pointed out, that the total mass, total electric charge, and total angular momentum are $M=M_{1}+M_{2}$, $Q=Q_{1}+Q_{2}$, and $J=J_{1}+J_{2}$, respectively. A peculiar characteristic of the binary system is that the seven physical parameters satisfy a dynamical law for interacting KN sources (BHs $\sigma_{i}^{2}\geq 0$ or naked singularities $\sigma_{i}^{2}< 0$) with struts defined by
\vspace{-0.1cm}
\be \mathfrak{q}\mathcal{P}_{0}-J_{1}\mathcal{P}_{2}-J_{2}\mathcal{P}_{1}=0. \label{condition1}\ee

\vspace{-0.1cm}
In this regard, the extreme limit solution is achieved by setting $\sigma_{1}=\sigma_{2}=0$ in Eq.\ (\ref{sigmas}), where this condition enables one the opportunity to express the angular momentum for each source in terms of the remaining parameters, thus getting
\vspace{-0.1cm}
\begin{align}
J_{1}&= \frac{\mathfrak{q} A_{0}B_{1}+ \varepsilon_{1}P_{0}\mathcal{P}_{1}\sqrt{P_{0}(R^{2}-\Delta)E_{0}d_{1}}}{G_{0}}, \quad \varepsilon_{1}=\pm 1, \nonu\\
J_{2}&= \frac{\mathfrak{q} A_{0}B_{2}+ \varepsilon_{2}P_{0}\mathcal{P}_{2}\sqrt{P_{0}(R^{2}-\Delta)E_{0}d_{2}}}{G_{0}}, \quad \varepsilon_{2}=\pm 1, \nonu\\
E_{0}&=4R\mathcal{P}_{1}+(R^{2}-\Delta)\Big[(P_{0}+2Q_{1}Q_{2})C_{1}
-2(R-\delta_{2})P_{0}H_{1+}\Big]\nonu\\
&\equiv 4R\mathcal{P}_{2}
+(R^{2}-\Delta)\Big[(P_{0}+2Q_{1}Q_{2})C_{2}-2(R+\delta_{2})P_{0}H_{2+}\Big],\nonu\\
d_{i}&=P_{0}^{2}D_{i}+\Bigg(\frac{\mathfrak{q} Q_{i}^{2}A_{0}}{\mathcal{P}_{i}}\Bigg)^{2}, \qquad \delta_{2}=M_{1}-M_{2},
\label{angularmomentum}\end{align}

\vspace{-0.1cm}
\noi where we have used the symmetric character of $G_{1}\equiv G_{2}=G_{0}$ and $A_{1}\equiv A_{2}=A_{0}$. On one hand, if one takes into account the case $\varepsilon_{1}=\varepsilon_{2}=\pm 1$, it might be possible to study co-rotating KN BHs, while on the other hand, the choice $\varepsilon_{1}=-\varepsilon_{2}=\pm 1$, permits the description of the corresponding counter-rotating scenario. The substitution of Eq.\ (\ref{angularmomentum}) into Eq.\ (\ref{condition1}) guides us to the simple formula
\vspace{-0.1cm}
\begin{align} &P_{0}(R^{2}-\Delta)E_{0} \Big(\sqrt{d_{2}}+ \epsilon \sqrt{d_{1}}\Big)^{2}
=\mathfrak{q}^{2}(E_{0}-2RA_{0})^{2},\quad
\epsilon=\pm 1, \label{cocounterrotating}\end{align}

\vspace{-0.1cm}
\noi and thereby the sign $+/-$ defines co/counter-rotating KN binary BHs. In order to illustrate how this dynamical law might be used to describe various scenarios among two interacting BHs, for instance, let us explore first the case of a binary system of unequal counterrotating KN BHs \cite{ICM2015} that arises immediately when $\epsilon=-1$ and $\mathfrak{q}=0$, where it is pretty much obvious that Eq.\ (\ref{cocounterrotating}) is satisfied with the condition
\vspace{-0.1cm}
\begin{align}
&\sigma_{1E}^{2}=\sigma_{2E}^{2}, \qquad
\sigma_{iE}=\sqrt{M_{i}^{2}- Q_{i}^{2}-2(-1)^{i}Q_{i} \frac{M_{2}Q_{1}-M_{1}Q_{2}}{R+M}},\quad i=1,2
\label{conditioncounter}\end{align}

\vspace{-0.1cm}
\noi where $\sigma_{iE}$ represents the BH horizons in electrostatic spacetimes \cite{VCH}. For such a case, both angular momenta displayed in Eq.\ (\ref{angularmomentum}) are reduced to
\vspace{-0.1cm}
\begin{align}
&J_{i}=\varepsilon_{i}\frac{\sigma_{iE}P_{1i}(R+M)}{\sqrt{P_{00}(R^{2}-\delta_{0})}}, \quad i=1,2, \quad \varepsilon_{1}=-\varepsilon_{2}=\pm 1,\nonu\\
P_{11}&=M_{1}\big[(R+M_{2})^{2}-M_{1}^{2}\big]-Q_{1}\big[Q_{2}R-\delta_{2}Q\big],\nonu\\
P_{00}&=\big[(R+M_{1})^{2}-M_{2}^{2}\big]\big[(R+M_{2})^{2}-M_{1}^{2}\big] 
+(\delta_{1}R+\delta_{2}Q)^{2}, \qquad
P_{12}=P_{11(1\leftrightarrow2)}, \nonu\\
\delta_{0}&=M^{2}-Q^{2}, \quad \delta_{1}=Q_{1}-Q_{2},
\label{momentacounter}\end{align}

\vspace{-0.1cm}
\noi and the substitution of these formulas inside of Eq.\ (\ref{simplePxandQx}) derives directly the simple results
\vspace{-0.1cm}
\begin{align}
q_{o}&=\frac{Q_{1}}{2} (R-2M_{2})-\frac{Q_{2}}{2}(R-2M_{1}),\qquad
b_{o}=\frac{Q_{2}P_{2}-Q_{1}P_{1}}{R+M},\nonu\\
P_{1}&=-\frac{J_{1}(R+\delta_{2})(R^{2}-\delta_{0})}{P_{11}}, \qquad P_{2}=\frac{J_{2}(R-\delta_{2})(R^{2}-\delta_{0})}{P_{12}}. \label{simpleqoboPi}\end{align}

\vspace{-0.1cm}
So, it is not quite difficult to show that Eqs.\ (\ref{noNUT}), (\ref{disconnect}), and (\ref{noBcharge}) are identically satisfied by the results given above in Eqs.\ (\ref{conditioncounter})-(\ref{simpleqoboPi}), when $\mathfrak{q}=0$ is considered.\footnote{In Ref.\ \cite{ICM2015} is contemplated a relation between the seven physical parameters that might be obtained from Eq.\ (\ref{condition1}), when $\mathfrak{q}=0$. It can be expressed in the very simple form $J_{1}P_{12}+J_{2}P_{11}=0$. Notice that this condition is satisfied if Eqs.\ (\ref{conditioncounter}) and (\ref{momentacounter}) are substituted into it.} On the other hand, the vacuum solution earlier studied in Ref.\ \cite{ICM2018} is the second trivial scenario that appears in the absence of electric charges $Q_{1}=Q_{2}=0$, where after choosing $\epsilon=+1$, Eq.\ (\ref{cocounterrotating}) provides us a bicubic equation for co-rotating Kerr BHs that is given by
\vspace{-0.1cm}
\begin{align}
&\Delta_{1} (4M_{1}M_{2}\mathfrak{q}^{2}-p_{1}p_{2})-4M_{1}M_{2}\mathfrak{q}^{2}R^{2}=0, \nonu\\
p_{1}&=(R+M_{1})^{2}-M_{2}^{2}+\mathfrak{q}^{2}, \qquad p_{2}=(R+M_{2})^{2}-M_{1}^{2}+\mathfrak{q}^{2}, \quad
\Delta_{1}=M^{2}-\mathfrak{q}^{2}.
\label{corotatingvacuum}\end{align}

\vspace{-0.1cm}
A trivial combination of Eqs.\ (\ref{simplePxandQx}), (\ref{angularmomentum}), and (\ref{corotatingvacuum}), will lead us to the following expressions in the co-rotating case
\vspace{-0.1cm}
\begin{align}
P_{1,2}&= \frac{(\Delta_{1}+MR)(M_{2}-M_{1})}{2\mathfrak{q}}\pm (M\mathfrak{q}-J), \nonu\\
J_{1}&=\frac{M_{1}^{2}\mathfrak{q}P_{0}}{M p_{1}}, \qquad J_{2}=\frac{M_{2}^{2}\mathfrak{q}P_{0}}{M p_{2}}.
\label{angularmomentumvacuumfinal}\end{align}

\vspace{-0.1cm}
In the meanwhile, whether $\epsilon=-1$ it is possible to derive another bicubic algebraic equation concerning counter-rotating Kerr BHs that now contains the form
\vspace{-0.1cm}
\be \Delta_{1} (4M_{1}M_{2}\mathfrak{q}^{2}-p_{1}p_{2})+4M_{1}M_{2}R^{2}(R+M)^{2}=0, \label{counterrotatingvacuum}\ee

\vspace{-0.1cm}
\noi where this binary configuration is entirely depicted by
\vspace{-0.1cm}
\begin{align}
P_{1,2}&= \frac{M\mathfrak{q}(R^{2}+MR-\Delta_{1})}{2(M_{2}-M_{1})(R+M)} \pm (M\mathfrak{q}-J), \nonu\\
J_{1}&=\frac{M_{1}^{2}\mathfrak{q}}{M_{2}-M_{1}}\Bigg( 1-\frac{2M_{2}(R+M)p_{2}}{4M_{1}M_{2}P_{0}-p_{1}p_{2}}\Bigg), \qquad
J_{2}=-\frac{M_{2}^{2}\mathfrak{q}}{M_{2}-M_{1}}\Bigg( 1-\frac{2M_{1}(R+M)p_{1}}{4M_{1}M_{2}P_{0}-p_{1}p_{2}}\Bigg).
\label{angularmomentumvacuumfinal}\end{align}

\vspace{-0.1cm}
Regarding now the most general case, where after non-trivial algebraic manipulations on Eqs.\ (\ref{simplePxandQx}), (\ref{angularmomentum}), and (\ref{cocounterrotating}), eventually we will get the algebraic set $\{q_{o},b_{o},P_{1},P_{2}\}$ that is giving us a complete description of extreme KN binary BHs in a physical representation; they read
\vspace{-0.1cm}
\begin{widetext}
\begin{align}
q_{o}&=\Bigg( \frac{Q_{1}C_{2}-Q_{2}C_{1}}{2}+M\Big[(M_{1}Q_{2}-M_{2}Q_{1})P_{0}
-Q_{1}Q_{2}(Q_{1}-Q_{2})(R+M)\Big]\Bigg)\frac{RP_{0}(R^{2}-\Delta)}{E_{0}-2RA_{0}},\nonu\\
b_{o}&=\Bigg[ \frac{Q_{1}}{\mathfrak{q}} \Bigg( RP_{0}\bigg( \frac{2H_{2+}(R^{2}-\Delta)+R(C_{2}+Q_{2}QP_{0})}{E_{0}-2RA_{0}}\bigg)-1\Bigg)
+\frac{Q_{2}}{\mathfrak{q}} \Bigg( RP_{0}\bigg( \frac{2H_{1+}(R^{2}-\Delta)+R(C_{1}+Q_{1}QP_{0})}{E_{0}-2RA_{0}}\bigg)-1\Bigg)\Bigg]\nonu\\
&\times \frac{(R^{2}-\Delta)}{2}, \nonu\\
P_{1,2}&=\frac{RP_{0}(d_{2}-d_{1})(R^{2}-\Delta)}{2\mathfrak{q}(E_{0}-2RA_{0})}\pm (M\mathfrak{q}-J),
\label{simplePxandQx2}\end{align}

\vspace{-0.1cm}
\noi while the total angular momentum can be expressed as follows
\vspace{-0.1cm}
\begin{align}
J&=M\mathfrak{q}-\Bigg(\frac{R+M}{\mathfrak{q}}-RP_{0}^{2}\frac{P_{0}(R^{2}-\Delta+MR)-2Q_{1}Q_{2}(R+M)}
{\mathfrak{q}(E_{0}-2RA_{0})}\Bigg)
 \frac{(R^{2}-\Delta)}{2}.
\label{totalmomenta}\end{align}
\end{widetext}

\vspace{-0.1cm}
Then we have that a whole description for co- and counter-rotating electrically charged BHs is made once the parameters contained in  Eqs.\ (\ref{simplePxandQx2}) and (\ref{totalmomenta}) are inserted inside the asymptotically flat exact solution, which is obtainable from Eqs.\ (\ref{Ernstextreme}) and (\ref{extreme}) by making simply $J_{0}=0$ and $B=0$, namely
\vspace{-0.1cm}
\begin{align} {\cal{E}}&=\frac{\Lambda-2\Gamma}{\Lambda+2\Gamma},\qquad \Phi=\frac{2 \chi}{\Lambda+2\Gamma}, \qquad f=\frac{\mathcal{D}}{\mathcal{N}},\qquad
\omega=\frac{R(y^{2}-1)(x-1)\big[(x+1)\Sigma\Pi-\Theta {\rm T}\big]}{2\mathcal{D}},\qquad e^{2\gamma}=\frac{\mathcal{D}}{R^{8}(x^{2}-y^{2})^{4}}, \nonu\\
\Lambda&=R^{2}\left[(R^{2}-\Delta)(x^{2}-y^{2})^{2}+\Delta(x^{4}-1)\right] +\Big\{|\mathfrak{p}|^{2}+\mathfrak{q} \mathfrak{r}-R^{2}(R^{2}-\Delta)\Big\}(y^{4}-1) \nonu\\
&+2iR \Big\{xy\Big[\big(\mathfrak{r}+\mathfrak{q}R^{2}\big)(y^{2}-1)-\mathfrak{q}R^{2}(x^{2}+y^{2}-2)\Big] -R{\rm s}_{1}\big(x^{2}+y^{2}-2x^{2}y^{2}\big) \Big\}, \nonu\\
\Gamma&=M\mathbb{P}_{1}-\varepsilon\mathbb{P}_{2}, \qquad
\chi= Q \mathbb{P}_{1}+2\mathfrak{q}_{o} \mathbb{P}_{2} \qquad
\mathbb{P}_{1}=R^{3} x(x^{2}-1)-(R \bar{\mathfrak{p}} x-i \mathfrak{r} y)(y^{2}-1), \nonu\\
\mathbb{P}_{2}&= R^{2}y (x^{2}-1)-\big[\mathfrak{p}y-i\mathfrak{q}Rx\big](y^{2}-1),\qquad
\mathcal{N}= \mathcal{D}+ \Theta \Pi-(1-y^{2})\Sigma {\rm T}, \qquad \mathcal{D}= \Theta^{2}+(x^{2}-1)(y^{2}-1)\Sigma^{2},\nonu\\
\Theta&=R^{2}\Big[(R^{2}-\Delta)(x^{2}-y^{2})^{2}+\Delta(x^{2}-1)^{2}\Big] + \Big[|\mathfrak{p}|^{2}+\mathfrak{q} \mathfrak{r}
-R^{2}(R^{2}-\Delta) \Big](y^{2}-1)^{2}, \nonu\\
\Sigma&=2R\Big(\mathfrak{q}R^{2}x^{2}-\mathfrak{r}y^{2} -2R{\rm s}_{1}xy\Big),\nonu\\
\Pi&= 4Rx\bigg\{MR^{2}(x^{2}-y^{2})
+(M\Delta+\mathfrak{q}{\rm s}_{2})(1+y^{2})+(2M^{2}-Q^{2})Rx  -2\mathfrak{q}{\rm s}_{1}y\bigg\} \nonu\\
&-4y\bigg\{ \mathfrak{b}_{o}\Big(R^{2}(x^{2}-y^{2})+\Delta(1+y^{2})+2MRx-2\mathfrak{b}_{o}y\Big) 
+{\rm s}_{1}{\rm s}_{2}(1+y^{2}) -2\big({\rm s}_{2}^{2}-2|\mathfrak{q}_{o}|^{2}\big)y  \bigg\}, \nonu\\
{\rm T}&=4\Big\{R\big[\mathfrak{a}_{o}+{\rm s}(Rx+M)+(\mathfrak{q}\mathfrak{b}_{o}
-M{\rm s}_{1})y\big](1+x)
+\big[M\mathfrak{r}+\mathfrak{b}_{o}{\rm{s}}_{1}+(R^{2}-\Delta_{o}){\rm{s}}_{2}\big](1-y^{2})\Big\},\nonu\\
\mathfrak{p}&=R^{2}-\Delta+i{\rm s}_{1},\quad \mathfrak{r}=2\mathfrak{a}_{o}-\mathfrak{q}(R^{2}-2\Delta), \quad \varepsilon=\mathfrak{b}_{o}+i{\rm{s}}_{2}, \quad
\mathfrak{a}_{o}=M{\rm s}_{2}+2b_{o}Q,  \quad
\mathfrak{b}_{o}=\big(\mathfrak{q}{\rm s}_{1}-2q_{o}Q\big)/M,
\label{extremeflat}\end{align}

\vspace{-0.1cm}
\noi where a physical representation in terms of the parameters $\{M_{1},M_{2},Q_{1},Q_{2},R\}$ will be achieved through the parameter $\mathfrak{q}$ once Eq.\ (\ref{cocounterrotating}) is taken into account. Finally, to complete the full solution, we show also the Kinnersley potential \cite{Kinnersley}
\vspace{-0.1cm}
\begin{align} \Phi_{2}&=\frac{(4\mathfrak{q}+iRxy)\chi-i \mathcal{I}}{\Lambda+2\Gamma}, \nonu\\
\mathcal{I}&=R^{2}\bigg\{ 2\mathfrak{q}_{o}\Big[Rx-4i\mathfrak{q}y-M(1-y^{2})\Big]-
Q\Big[\bar{\varepsilon}(1+y^{2}) +\big(2MRx-4\mathfrak{b}_{o}y-2\bar{p}+R^{2} \big)y +2i\mathfrak{q}Rx\Big]\bigg\}(x^{2}-1)\nonu\\
&+\bigg\{2\mathfrak{q}_{o} \Big[\left(Mp-i\mathfrak{q}\bar{\varepsilon}\right)(1+y^{2})-Rx \big(4MRx-2\bar{\varepsilon}y+\bar{p}+4\Delta\big) 
+ i(2\mathfrak{q}R^{2}-r)y\Big]\nonu \\
&-Q\Big[(\bar{\varepsilon}\bar{p}-iMr)(1+y^{2})-Rx\big[2\varepsilon R x+ i(2r-\mathfrak{q}R^{2}) \big]
-\Big(R^{2}(\bar{p}+2R^{2}-2\Delta)-2(|p|^{2}+\mathfrak{q}r)\Big)y\Big] \bigg\}(y^{2}-1),
\label{Kinnersley}\end{align}

\vspace{-0.1cm}
\noi with the aim to derive the magnetic potential by considering its real part. This potential is useful to consider the contribution of the Dirac string in the horizon mass \cite{Tomi,Galtsov,ICM2020}.

\vspace{-0.2cm}
\subsection{Physical characteristics of extreme KN binary BHs}
\vspace{-0.3cm}
The thermodynamical properties of each extreme KN BH satisfy the well-known Smarr formula for the mass \cite{Tomi,Smarr}
\vspace{-0.1cm}
\begin{align}
M_{i}&=2\Omega_{i}J_{i}+ \Phi_{i}^{H}Q_{i}, \qquad i=1,2, \label{Massformula}\end{align}

\vspace{-0.1cm}
\noi where $\Omega_{i}$ and $\Phi_{i}$ define the angular velocity and electric potential in the corotating frame of each BH. Another, important thermodynamical aspect to be considered in the binary system is the area of the horizon $S_{i}$. In the extreme limit case of BHs, their corresponding formulas are obtained by setting $\sigma_{i}=0$ in Eq.\ (33) of Ref.\ \cite{ICM2021}. The result is
\vspace{-0.1cm}
\begin{align}
\Omega_{i}&= \frac{\mathfrak{q} A_{i}}{P_{0}\mathcal{P}_{i}} + \frac{J_{i}P_{0}^{3}\mathcal{P}_{i}R^{2}(R^{2}-\Delta)
}{\mathcal{P}_{i}^{2}\mathcal{N}_{i}^{2}+
P_{0}^{2}(R^{2}-\Delta)^{2}\mathcal{M}_{i}^{2}}, \qquad
\Phi_{i}^{H}=\frac{M_{i}-2\Omega_{i}J_{i}}{Q_{i}},\qquad
S_{i}=4\pi \frac{\mathcal{P}_{i}^{2}\mathcal{N}_{i}^{2}+
P_{0}^{2}(R^{2}-\Delta)^{2}\mathcal{M}_{i}^{2}}{R^{2}P_{0}\mathcal{P}_{i}^{2}},\nonu\\
\mathcal{N}_{i}&=M_{i} P_{0}-2\mathfrak{q}J_{i}-Q_{i}Q(R+M),\qquad
\mathcal{M}_{i}=J_{i}C_{i}+\mathfrak{q}Q_{i}^{2}\left[M_{i}P_{0}+
Q_{1}Q_{2}(R+M)\right], \quad i=1,2. \label{Horizonproperties}\end{align}

\vspace{-0.1cm}
On the other hand, the interaction force related to the strut has the form \cite{ICM2021}
\vspace{-0.1cm}
\begin{align}
\mathcal{F}&=\frac{\mathcal{N}_{0}}{P_{0}^{3}(R^{2}-M^{2}+Q^{2}+\mathfrak{q}^{2})},  \nonu\\
\mathcal{N}_{0}&=(M_{1}M_{2}P_{0}^{2}-\mathfrak{q}^{2}Q_{1}^{2}Q_{2}^{2})\left[(R+M)^{2}
-\mathfrak{q}^{2}\right]-(Q_{1}-F_{0})(Q_{2}+F_{0})P_{0}^{3}
+\mathfrak{q}^{2}\Big\{(M_{1}Q_{2}-M_{2}Q_{1})^{2}P_{0}\nonu\\
+&Q_{1}Q_{2}\left[ 2(R^{2}+MR+\mathfrak{q}^{2})P_{0}+(P_{0}+Q_{1}Q_{2})Q^{2}\right] \Big\},
\label{force} \end{align}

\vspace{-0.1cm}
\noi and it acquires the same aspect regardless if the binary system is co/counter-rotating. The distinction between each configuration will be dictated by the choice of the sign $+/-$ in the dynamical law shown in Eq.\ (\ref{cocounterrotating}). The conical singularity in the middle region among the BHs can be removed if the force equals zero. However, a non vanishing force can give us more insight on how the sources are attracting or repelling each other via the gravitational, spin-spin, and electric interactions. Besides, the force provides the limits of the interaction distance, for instance, the merger limit of BHs is obtainable by equating the denominator of its formula to zero, thus getting the value $R_{0}=\sqrt{M^{2}-Q^{2}-\mathfrak{q}^{2}}$, where $\mathfrak{q}=J/M$ [see Eq.\ (\ref{totalmomenta})]. Luckily, the merger limit of extreme BHs brings us rather simple expressions for $\Omega_{i}$ and $\Phi_{i}$, which are given by
\vspace{-0.1cm}
\begin{align} \Omega_{1}&=\Omega_{2}=\frac{J/M}{d_{0}}, \qquad
\Phi_{i}=\frac{Q(R_{0}+M)}{d_{0}}+ \frac{R_{0}}{2Q_{i}}, \quad i=1,2, \nonu\\
d_{0}&=(R_{0}+M)^{2}+ (J/M)^{2}.\end{align}

\vspace{-0.1cm}
\noi while each angular momentum takes the final form
\vspace{-0.1cm}
\begin{align} J_{1}&=M_{1} \mathfrak{q} +\frac{Q\nu}
{2\mathfrak{q}}+R_{0} \frac{R\big[2\delta_{2}(R_{0}+M)-Q\delta_{1}\big]}{4\mathfrak{q}}, \qquad
J_{2}=M_{2}\mathfrak{q}-\frac{Q\nu}
{2\mathfrak{q}}-R_{0} \frac{R\big[2\delta_{2}(R_{0}+M)-Q\delta_{1}\big]}{4\mathfrak{q}}, \nonu\\
\nu&=M_{1}Q_{2}-M_{2}Q_{1}. \label{extremerelations0}\end{align}

\vspace{-0.1cm}
On one hand, in the co-rotating case, $\mathfrak{q}=\sqrt{M^{2}-Q^{2}}$ at $R_{0}=0$, where it can be possible to recover the well-known expression for extreme KN BHs
\vspace{-0.1cm}
\be J_{1}+J_{2}=(M_{1}+M_{2})\sqrt{(M_{1}+M_{2})^{2}-(Q_{1}+Q_{2})^{2}}. \label{extremerelation} \ee

\vspace{-0.1cm}
On the other hand, the counter-rotating case establishes as $\mathfrak{q}=0$ the lowest value for $\mathfrak{q}$ at distance $R_{0}=\sqrt{M^{2}-Q^{2}}$. Naturally that, in this physical scenario, the total angular momentum of the system is $J=0$, where both angular momenta fulfill the condition $J_{1}=-J_{2}=\infty$. Moreover, when the BHs are far away from each other, in the limit $R \rightarrow \infty$, the parameter $\mathfrak{q} \rightarrow  J_{1}/M_{1}+\epsilon J_{2}/M_{2}$, $\epsilon=\pm 1$, where now each angular momentum satisfies the relation $J_{i}=M_{i}\sqrt{M_{i}^{2}-Q_{i}^{2}}$ for extreme KN BHs. Once again, it must be recalled that the sign $+/-$ for $\epsilon$ is related to co/counter-rotating KN BHs, respectively.

Continuing with the analysis, but centering our attention on the co-rotating scenario, one should be aware that the interaction force as well as the area of the horizon become indeterminate at the value $R=0$. To calculate their correct expressions, one must apply a Taylor expansion around $R=0$ after using the term $\mathfrak{q}=\sqrt{M^{2}-Q^{2}}
+C_{0}R$ in Eq.\ (\ref{cocounterrotating}) with the objective to obtain the correct value for $C_{0}$, which is governed by the quadratic equation
\vspace{-0.1cm}
\begin{align} & 4C_{0}^{2}+4 \sqrt{(\alpha_{+}^{2}-1)(1-\alpha_{-}^{2})} C_{0}+\alpha_{-}^{2}-1=0, \nonu \\
\alpha_{\pm}&=2\sqrt{\frac{2M^{2}-Q^{2}}{\alpha_{0}}}\Bigg(\sqrt{4J_{1}^{2}+Q_{1}^{4}} \pm \sqrt{4J_{2}^{2}+Q_{2}^{4}}\Bigg), \nonu\\
\alpha_{0}&= \delta_{0}(4M^{2}-Q^{2}-\delta_{1}^{2})^{2}+4(2M^{2}-Q^{2})^{2}\delta_{2}^{2},
\label{quadratic}\end{align}

\vspace{-0.1cm}
\noi and, therefore, it can be shown that the final expressions for the area of the horizon and the force are given by
\vspace{-0.1cm}
\begin{align}
S_{i}&=\frac{4\pi(2M^{2}-Q^{2})^{3/2} \sqrt{4J_{i}^{2}+Q_{i}^{4}}}{\sqrt{\alpha_{0}} \alpha_{+}}\Bigg[1+ \bigg(\sqrt{\alpha_{+}^{2}-1}-\epsilon_{0} \alpha_{+}\bigg)^{2}\Bigg], \nonu\\
\mathcal{F}&=\frac{\alpha_{0}\Big(\sqrt{\alpha_{+}^{2}-1}+\epsilon_{0} \alpha_{+}\Big)^{2}-4(2M^{2}-Q^{2})^{3}}{16 (2M^{2}-Q^{2})^{3}}, \quad i=1,2,
 \label{mergerformulas}
\end{align}

\vspace{-0.1cm}
\noi where $\epsilon_{0}= \pm 1$. In this occasion, the signs $+/-$ define two different scenarios during the touching limit where both sources attract/repel to each other. In order to verify the accuracy of our result, for $Q_{1}=Q_{2}=0$, it is possible to attain directly from Eq.\ (\ref{mergerformulas}) the result \cite{CCHV},
\vspace{-0.1cm}
\begin{align} S_{i}&=8\pi M_{i}(M_{1}+M_{2})^{2}\bigg(\frac{M_{1}+M_{2}-\epsilon_{0} \sqrt{2 M_{1}M_{2}}}{M_{1}^{2}+M_{2}^{2}}\bigg), \qquad
\mathcal{F}=\frac{M_{1}M_{2}+ \epsilon_{0} \sqrt{ 2M_{1}M_{2}}(M_{1}+M_{2})}{2(M_{1}+M_{2})^{2}}, \quad i=1,2. \end{align}

\vspace{-0.1cm}
\noi where the attractive scenario has been deduced before in Ref.\ \cite{CiafreII}, but it was expressed in terms of dimensionless parameters. Finally, after performing a first order expansion around $R=0$ and the simple coordinate changes $\rho=(r-M)\sin \theta$,  $z=(r-M)\cos \theta$ on Eq.\ (\ref{extremeflat}), it is possible to get
\vspace{-0.1cm}
\begin{align}
&\hspace{-0.8cm} {\cal{E}} \simeq 1-\frac{2M}{r-i a\cos \theta}\Bigg(1-\mathcal{K} \frac{(r-M)\cos\theta-i a \sin^{2} \theta}{r-ia \cos \theta} \frac{R}{r-M}\Bigg), \quad
\Phi \simeq \frac{Q}{r-i a\cos \theta}\Bigg(1-\mathcal{K} \frac{(r-M)\cos\theta-i a \sin^{2} \theta}{r-ia \cos \theta} \frac{R}{r-M}\Bigg),\nonu\\
f&\simeq f_{0}\Bigg[1+2\mathcal{K}\frac{(Mr+a^{2})\Xi-(2Mr-Q^{2})(Mr+a^{2}\cos 2\theta)}{\Xi(\Xi-2Mr+Q^{2})}\frac{R\cos\theta}{r-M}\Bigg], \qquad e^{2\gamma}\simeq e^{2\gamma_{0}}+\frac{4a^{2}\mathcal{K}\sin^{2}\theta}{(r-M)^{3}}R\cos \theta,\nonu\\
\omega&\simeq \omega_{0}
\Bigg[1-2\mathcal{K}\Bigg(\frac{2(r-M)}{\Xi-2Mr+Q^{2}}-\frac{M}{2Mr-Q^{2}}\Bigg)R \cos \theta \Bigg], \nonu\\
\mathcal{K}&=\frac{(2M^{2}-Q^{2})\big[Q\delta_{1}(4M^{2}-Q^{2}-\delta^{2}_{1})
-4M(2M^{2}-Q^{2})\delta_{2}\big]}
{\alpha_{0}}\Bigg(1-\frac{\epsilon_{0}\sqrt{\alpha_{+}^{2}-1}}{\alpha_{+}}\Bigg), \nonu\\ 
f_{0}&=1-\frac{2Mr-Q^{2}}{\Xi}, \quad
e^{2\gamma_{0}}=1-\frac{a^{2}\sin^{2} \theta}{(r-M)^{2}}, \quad  \omega_{0}=-\frac{a(2Mr-Q^{2}) \sin^{2} \theta}{\Xi-2Mr+Q^{2}},\qquad
\Xi=r^{2}+a^{2}\cos^{2} \theta, \quad \epsilon_{0}=\pm 1, \label{Ernstnearhorizon}\end{align}

\vspace{-0.1cm}
\noi being $a=J/M$ the angular momentum per unit mass and $(r,\theta)$ are the Boyer-Lindquist coordinates. Eq.\ (\ref{Ernstnearhorizon}) defines a deformed metric for a near horizon extreme binary KN BH, where in the physical limit at $R=0$ it is  possible to recover the metric for a single extreme KN BH of mass $M=M_{1}+M_{2}$, electric charge $Q=Q_{1}+Q_{2}$, and angular momentum $J=M \sqrt{M^{2}-Q^{2}}$, in other words
\vspace{-0.1cm}
\begin{align} ds^{2}&=f_{0}^{-1}\Big[e^{2\gamma_{0}}\big[dr^{2}+(r-M)^{2}d\theta^{2}\big]+(r-M)^{2}\sin^{2} \theta d\varphi^{2}\Big]- f_{0}(dt-\omega_{0} d\varphi)^{2}.
\label{KerrNewmanmetric} \end{align}

\vspace{-0.6cm}
\section{Conclusion}
\vspace{-0.3cm}
The derivation of the metric that completely characterizes unequal configurations of extreme KN BHs in a physical representation finally has been succeeded. The task of solving the conditions on the axis and the one eliminating the magnetic charges is accomplished by adopting a fitting parametrization that has been earlier introduced in \cite{ICM2021}. It follows that the asymptotically flat metric has been written in a quite simple form by means of
Perjes' approach \cite{Perjes}, where it contains a physical representation in terms of the set $\{M_{1},M_{2},Q_{1},Q_{2},R\}$ that is very suitable to concrete some applications in rotating charged binary systems. Similarly to the non-extreme case \cite{ICM2021}, the physical parameters are related to each other through an algebraic equation that represents a dynamical law for interacting BHs with struts. Unfortunately, there is no chance to solve exactly this higher degree equation except for some special unequal cases \cite{ICM2018,ICM2015}.

Due to the fact that our solution reported in this work is presented with a more physical aspect, the physical limits of the interaction of BHs can be readily identified in both co- and counter-rotating cases. Even better, the thermodynamical characteristics of each BH during the merger limit have been also derived and concisely introduced.   With regard to the co-rotating case, in particular, during the merger limit of binary BHs, it is possible to conceive an attractive or repulsive final state. In this respect, the deformed metric for a near horizon extreme binary KN BH is also obtained, from which we do not exclude that it might be helpful to develop some analytical studies related to the collision of two BHs like gravitational waves by assuming a quasi stationary process, in a similar way to that previously considered in \cite{JSN}.

\vspace{-0.4cm}
\section*{Acknowledgements}
\vspace{-0.3cm}
The author acknowledges the financial support of SNI-CONACyT, M\'exico, grant with CVU No. 173252.

\end{document}